\documentclass[12pt, draftclsnofoot, onecolumn]{IEEEtran}

\usepackage[margin=1in,a4paper]{geometry}
\usepackage{amssymb}
\usepackage{amsthm}
\usepackage{array}
\usepackage{tabularx}

\usepackage{amsmath}
\usepackage{graphicx}

\usepackage{breqn}
\newtheorem{prop}{Proposition}

\usepackage{algorithm}
\usepackage{algpseudocode}

\usepackage[colorlinks=true, linkcolor=red, urlcolor=blue, citecolor=gray]{hyperref}
\usepackage{xcolor}

\begin{document}



\title{Energy optimization for Full-Duplex Wireless-Powered IoT Networks using Rotary-Wing UAV with Multiple Antennas}

\author{Leyla Fathollahi, Mahmood Mohassel Feghhi\footnote{Leyla Fathollahi and Mahmood Mohassel Feghhi are with the Faculty of Electrical and Computer Engineering, University of Tabriz, Tabriz, Iran (Email: L\_fathollahi98@ms.tabrizu.ac, mohasselfeghhi@tabrizu.ac.ir).}, Mahmoud Atashbar\footnote{Mahmoud Atashbar is with the Department of Electrical Engineering, Azarbaijan Shahid Madani University, Tabriz, Iran (Email: atashbar@azaruniv.ac.ir).}}

\maketitle

\begin{abstract}
In this paper, we propose a novel design for the rotary-wing unmanned aerial vehicle (UAV)-enabled full-duplex (FD) wireless-powered Internet of Things (IoT) networks. In this network, the UAV is equipped with an antenna array, and the $K$ IoT sensors, which are distributed randomly, use single-antenna to communicate. By sending the energy, the UAV as a hybrid access point, charges the sensors and collects information from them. Then, to manage the time and optimize the energy, the sensors are divided into N groups, so that the UAV equipped with multi-input multi-output (MIMO) technology can serve the sensors in a group, during the total time $T$. We provide a simple implementation of the wireless power transfer protocol in the sensors by using the time division multiple access (TDMA) scheme to receive information from the users. In other words, the sensors of each group receive energy from the UAV, when it hovers over the sensors of the previous group, and also when the UAV flies over the previous group to the current group. The sensors of each group send their information to the UAV, when the UAV is hovering over their group. Under these assumptions, we formulate two optimization problems: a sum throughput maximization problem, and a total time minimization problem. Numerical results show that our proposed optimal network provides better performance than the existing networks. In fact, our novel design can serve more sensors at the cost of using more antennas compared to that of the conventional networks. 
\end{abstract}

%\begin{keyword}
%%% keywords here, in the form: keyword \sep keyword
%Internet of things \sep unmanned aerial vehicle  \sep multi-input multi-output \sep wireless power transfer protocol \sep convex optimization.
%
%\end{keyword}


\section{Introduction}
Due to the quick expansion of the use of Internet of Things (IoT), this technology is ready to become one of the key requirements in the future. The structure of the IoT allows the devices in the network to connect to each other or to the central processor and send information to them. This technology can be used in distant  areas or regions, such as transportation (sea, road, rail, air, …), navy management, logistics, solar energy, oil and gas extraction, smart measuring tools, agriculture, environmental monitoring and mining, especially the areas which are not technically or economically accessible \cite{1}.

The widespread use of the IoT has turned energy consumption management into a big challenge. These devices, which are usually compact, use small batteries such as coin cells with a life of approximately two years \cite{2}. Replacing or recharging these batteries is inconvenient due to their location in remote areas or economic issues. Therefore, IoT industry researchers decided to use wireless sensors. These devices do not need to follow a specific pattern in placement, and the network designer can lead the network more flexibly.

An electromagnetic field is created by the combination of electric and magnetic fields. In the frequency range of electromagnetic waves, the wavelength is reduced and wireless transmission of power over long distances is possible. The use of a technology such as radio frequency (RF) energy transferring, which can charge devices with a fixed or mobile location from a distance of tens of kilometers, is promising \cite{3}. In recent years, the wireless power transfer (WPT) technology, which means supplying the required power to a device without physical contact with the source, has attracted the attention of many researchers \cite{4}. The IoT networks, with long-range communication and low-power consumption ability, are generally composed of the ground users and the hybrid access points (HAPs). The HAP sends energy to the sensors and receives the information from them for further processing. On the other hand, the sensors send the required data to the HAP, using the received energy, via the energy harvesting (EH) protocol. In conventional networks, a fixed HAP serves the sensors; however, it is challenging in large networks, when the ground users are placed sparsely. In fact, the sensors, which are located far from the energy source, receive less energy. The solution is to increase the number of these energy sources, which is not economical. After a plenty of research, the researchers came to the conclusion that moving HAPs can provide more stable and reliable energy for sensors, due to their flexibility in movement \cite{5}.

Due to the unique characteristics of UAVs, eventually replacing them as mobile HAPs with traditional fixed HAPs has created a huge revolution in wireless communication networks. Recently, the WPT using UAV is a hopeful technology to prepare stable energy for low-power sensors in large-scale networks, thanks to the movability, flexibility, maneuverability, simplicity of positioning, and relatively low cost. Flexible flight allows the energy transmitters placed on the UAV to charge ground devices more effectively and increase the capacity of existing cellular systems. In contrast to typical ground base stations, the UAV as an aerial base station has many capabilities, such as height controlling, Ability to the line of sight (LOS) communication with the sensors, and avoiding of the obstacles \cite{6}.

The security of the communication among the UAV and the sensors in the network is one of the most fundamental issues, so that the UAV can perform its tasks without leaking sensitive information to unknown users. In \cite{7}, the user anonymity and the mutual authentication are analyzed as security parameters against attackers. Efficient 3D deployment and the UAV trajectory optimization, as the most important issues in the UAV-equipped networks, have been investigated in \cite{8}. It is a function of the network placement environment, the location of the sensors, the flight altitude and the characteristics of the channel among the UAV and the sensors. Evaluating network performance to analyze the impact of design parameters is another fundamental issue for designing UAV-based communication systems. The UAVs as independent aerial base stations can be used for improvement of the downlink coverage and rate performance of a device-to-device communication system, co-exists with a single UAV, as it is done in \cite{9}.

The flexibility, the LOS communication, the energy limitations and the backhaul connectivity in network planning for a UAV-based network are some challenging issues in the field of wireless communications. A new method for optimizing the joint deployment of cellular base stations and backhaul links as a multi-objective optimization problem has been presented in \cite{10}, considering the backhaul capacity and the outage probability as constraints.

The research in \cite{11} is about a communication network equipped with a UAV with rotary-wing. In this system, the target is to minimize the consumption of the UAV driving energy and the communication energy, while ensuring that the ground sensors receive enough energy to perform the operations. In \cite{12}, a technique for energy efficiency to improve the network lifetime has been discussed. This paper tries to minimize the maximum energy consumption of all sensors by optimizing the sensors' wake up schedule and the path of the UAV. In \cite{13}, a set of energy receivers in known locations are served by the UAV as an energy transmitter. The optimal solution to transfer the maximum energy to the users is obtained, when the UAV hovers in one place during the entire charging period. 

In this paper, we focus on a new design for the rotary-wing UAV-enabled FD wireless-powered IoT networks. In this wide network, $K$ single-antenna sensors with sparse distribution are served by a UAV with an antenna array. Under these assumptions, in order to optimize the total operation time and energy consumption of the UAV, we divide the sensors into $N$ groups. In our scheme, each group of sensors receives energy from the UAV, when the UAV is hovering over the previous sensor group, and also when the UAV flies over the previous sensor group to the current sensor group. Each group of sensors sends their information to the UAV, when the UAV is hovering over them. As a result, unlike \cite{14}, the UAV serves a group of sensors instead of one sensor each time the UAV flies and hovers. Unlike \cite{20}, obviously, if the UAV transmits wireless power to the sensors of the next group only when it is flying between two groups, it will cause a waste of time resources. The contributions of this paper are as follows:
\begin{itemize}
\item Unlike the works done in \cite{15}, \cite{16}, \cite{17} and \cite{18}, the UAV in our scheme has more than one antenna to simultaneously transfer energy to sensors and receive information from them. Compared to the UAV in \cite{14}, the UAV, in our proposed system model, has more antennas to receive information from the sensors to increase the information transmission rate and optimize the flight and hovering time.
\item In contrast to \cite{19}, the causality of energy is also considered in our method. That is, each sensor can only use energy to transmit information to the UAV, which is received from the UAV in the previous time frame.
\item In the research conducted in \cite{20}, the flight speed of the UAV was not considered; however, the maximum speed is one of the limitations of the optimization problems, in our research.
\item Due to the real needs of the receiver, the limitation of WPT has also been one of the main criteria in our scheme. A case not considered in \cite{20}.
\end{itemize}

The rest of the paper is as follows: Section~\ref{sec:Sys_Struc} describes a rotary-wing UAV-enabled FD wireless-powered IoT networks using the antenna array in the UAV. In Section~\ref{sec:STM} and ~\ref{sec:TTM}, we formulate the sum throughput maximization (STM) and the total time minimization (TTM) problems and describe their optimal solution in closed forms. In section~\ref{sec:simulations}, we obtain the numerical results using proposed generalized algorithms, and compare them with the simulation results of the previous algorithms. Section~\ref{sec:Conclusion} draws some conclusions.

\section{System Structure}\label{sec:Sys_Struc}
\begin{figure}
\centering{\includegraphics[scale=0.35]{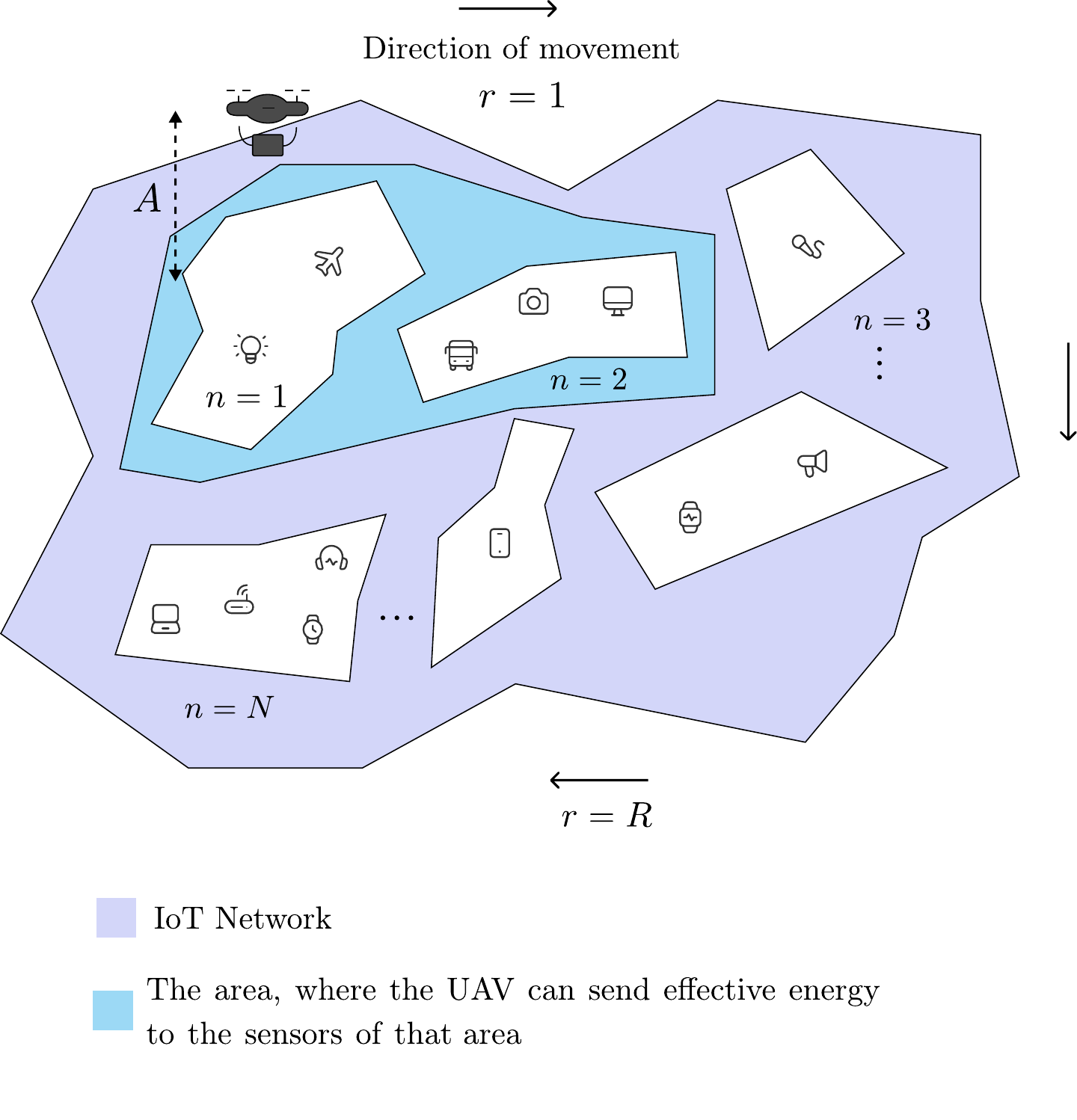}}
\caption{Proposed IoT system equipped with UAV.}\label{Fig1}
\end{figure}

According to Figure~\ref{Fig1}, we propose a new scheme for the UAV-enabled FD wireless-powered IoT networks using the antenna array in the UAV. In this network, $K$ single-antenna ground users are sparsely located in the network. Also, we use uniform linear array (ULA) to arrange the antennas in the UAV. A ULA is a set of antenna elements, which are arranged along a straight line with equal distances from each other. The purpose of using this type of array is to enhance the signal to noise ratio in a specific direction \cite{21}.

The UAV sends the required energy to the sensors using WPT and collects information from them. In contrast, the sensors receive energy from the UAV and send their information to it, using wireless information transfer (WIT) protocol. Due to the limitation in time resources, it is important to use effective methods for time management during the operations. For this reason, the sensors in the network can be divided into $N$ groups of sensors. A UAV equipped with MIMO technology collects information from the sensors in a group by several antennas during the total time $T$ and sends energy to them with other antennas. In this case, the power gain of the channel is different in the downlink and the uplink channels.
Also, the primary antennas of the UAV are installed with the purpose of providing energy and other antennas in the antenna array to collect information from the sensors. Assuming that the ground users are sparsely distributed in the IoT network and are not mobile, we consider the flight height of the UAV is constant. The changes in the UAV altitude may have effects on the performance of WPT and WIT. The movement of the UAV in a fixed path from the beginning to the end of the network causes the sensors, which are far away from the path of the UAV, to receive less energy. Therefore, we can consider the vertical and horizontal paths in the entire network, so that the UAV can serve all the sensors effectively. In the following three subsections, we analyze the proposed network based on mathematical formulas.

\subsection{Analysis of the placement of network elements}
The location of the sensors in the network and the image of the location of the UAV antennas in two-dimensional (2D) coordinates can be expressed by \eqref{eq:1} and \eqref{eq:2}, respectively: 
\begin{equation}
	{{\bf{w}}_{\bf{i}}} = {[{x_i},{y_i}]^T},\,\forall i \in \mathcal{K} \buildrel \Delta \over = \{ 1,2,...,K\} \label{eq:1}
\end{equation}

\begin{equation}
	{{\bf{q}}_k}{\bf{(t)}} = {[{x_k}(t),{y_k}(t)]^T}\,,\forall k \in \mathcal{M} \buildrel \Delta \over = \{ 1,2,...,M\}  \label{eq:2}
\end{equation}

According to \eqref{eq:1} and \eqref{eq:2}, the communication distance among the UAV antennas and the sensors at any moment $t$ can be shown as 
 ${d_{ki}}(t) = \sqrt {{{\left( {{L_{ki}}(t)} \right)}^2} + {A^2}},$
where ${L_{ki}}(t) = \left\| {{q_k}(t) - \left. {{w_i}} \right\|} \right.$ is the horizontal distance between the image of the location of the $k$-\textit{th} UAV antenna and the $i$-\textit{th} sensor at time $t$. In our scheme, we assume that the amount of wireless power transmission is bounded, and the maximum distance of wireless power transmission between the UAV and the sensor is determined by $d_{\mathrm{max}}$. This means, outside the range of $d_{\mathrm{max}}$, the signal power is ineffective for energy harvesting operation. Therefore, the maximum horizontal distance among the antennas in the UAV antenna array and the sensors is also limited and is calculated as
${L_{\max }} = \sqrt {d_{\max }^2 - {A^2}} $.
In order to transmit reliable information among the UAV and the sensors, we are looking for optimal points. The UAV stops at optimal points and hovers over a group of sensors. In this case, the closest distance between the UAV and the sensors of the desired group is established. Considering the existence of $N$ groups of sensors, ${n} \in \aleph  = \{ 1,2,...,N\} $ and $S_n$ as the set of all sensors belonging to the $n$-\textit{th} group, it can be said that the distance between the sensors of the $n$-\textit{th} and $(n-1)$-\textit{th} groups is equal to $\mathrm{dis}_n(t)$ . This distance is defined as 
$\mathrm{dis}_n(t) = [{q_1}(t)\left| {i \in {S_n}} \right.] - [{q_1}(t)\left| {i \in {S_{n - 1}}} \right.]$, which must satisfy the two conditions $\mathrm{dis}_n(t) \le {d_{\max }}$ and $\mathrm{dis}_n(t) + \mathrm{dis}_{n+1}(t) > {d_{\max }}$.

\begin{figure}
\centering{\includegraphics[scale=0.45]{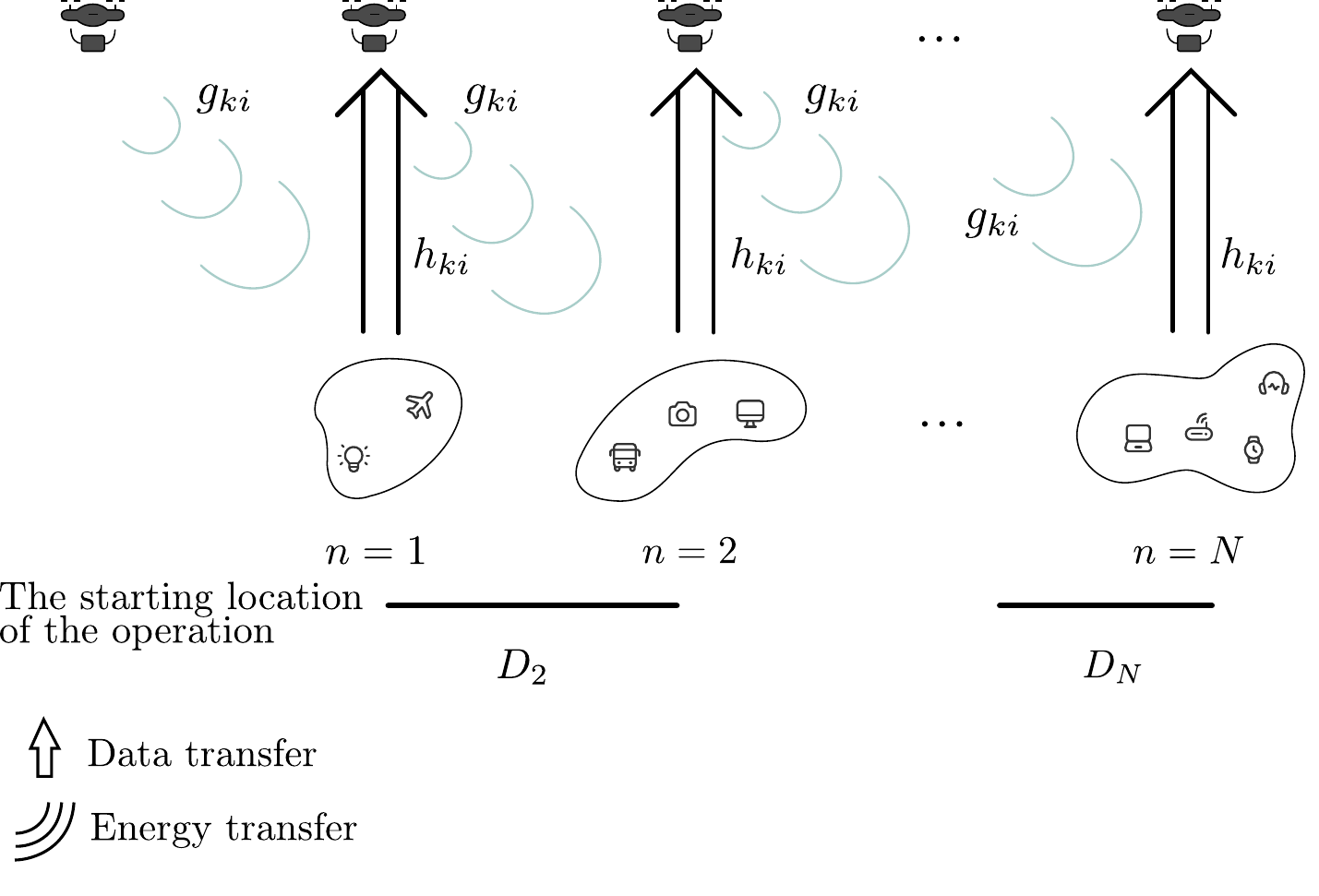}}
\caption{1D view of information transmission and energy harvesting in the proposed network.}\label{Fig2}
\end{figure}

Due to the sparsly distribution of the sensors, the one-dimensional (1D) model, instead of 2D model of the network is shown in Figure~\ref{Fig2}. The UAV hovers over the sensors of the $n$-\textit{th} group to collect information, while simultaneously charging the sensors of the $(n+1)$-\textit{th} group. The hovering time of the UAV on the sensors of the $n$-\textit{th} group is indicated by $\tau_n$. Here $\tau_0$ is the hovering time of the UAV at the starting location of the operation. The UAV can collect information from the sensors only when hovering over a group, so the time to collect information from $n$-\textit{th} group is equal to $\tau_n$. Also, $\zeta_n$ is the flight time of the UAV from the sensors of the $(n-1)$-\textit{th} group to the sensors of the $n$-\textit{th} group.
One of the most important factors for  the UAV operations is the total flight time due to limited energy resources. Therefore, we use a FD technology to simultaneously perform data collection and wireless power transmission, which optimizes the use of limited time resources.

In the following, according to the drawn network scheme, we assume that the UAV moves in the several horizontal paths, so that it can serve all available sensors effectively. Therefore, the UAV movement paths on different groups of sensors are $r \in \Re  \buildrel \Delta \over = \{ 1,2,...,R\} $. Also, we consider $\mathbb{N}_r$ as the set of all existing odd or even horizontal paths. In general, the coordinates of the reference antenna of the antenna array on the UAV in the time interval of movement from the sensors of the $(n-1)$-\textit{th} group to the sensors of the $n$-\textit{th} group will be equal to the following relations:

\begin{equation}
	x(t) = {v_n} t + {x_{n - 1}} \label{eq:3}
\end{equation}
\begin{equation} 
	y(t) = \tilde{y}, \label{eq:4}
\end{equation}
where $\tilde{y}$ is the vertical component of the desired movement path. If we consider $D_n$ as the distance between the stopping points between the $(n-1)$-\textit{th} group and the $n$-\textit{th} group, the average speed of the UAV while flying from the $(n-1)$-\textit{th} group to the $n$-\textit{th} group will be equal to $\dfrac{D_n}{\zeta_n}$.
So $\left[x_n,y_n\right]$ can be defined as the coordinates of the reference antenna of the antenna array on the UAV as follows:
\begin{equation}
	x_n = v_n \zeta_n + x_{n-1}  \label{eq:5}
\end{equation}
\begin{equation}
	y_n = \tilde{y}, \label{eq:6}
\end{equation}
The movement of the UAV is always perpendicular to the direction of movement, so the direction of the movement of the UAV and the direction of the antenna array on the UAV will be perpendicular to each other. In this case, $L_{ki}^{(n)}$  can be rewritten as follows:
\begin{equation}
	L_{ki}^{(n)} = {\sqrt {{{({v_n}{\zeta _n} + {x_{n - 1}} - {x_i})}^2} + {{({y^'} + (k - 1)\delta  - {y_i})}^2}}},     
\end{equation}
where $\delta$ is the fixed distance among the UAV antennas. 

\subsection{Channel Model}
In this system, we call the transmission constant power $P_t$ and consider the self-interference cancellation from the transmitter antenna to the receiver antenna. The digital and analog interferences, as self-interference cancellation, are negligible and at the noise level. This will be accurate, when the receiver antenna has a true estimate of the transmitted signal. In addition, beamforming is also useful for WPT. The UAV flies quickly during the operation and the channel among the UAV and the sensors is constantly changing, and the beamforming technology usually needs real information about the state of this channel. Here, the environment is rural and the target channel among the sensors and the UAV is considered without obstacles such as tall buildings. Therefore, we assume that the channel among the sensors and the UAV is the LOS wireless channel. As a result, the path loss will be $\dfrac{1}{d^2}$ , which is widely used in UAV-equipped wireless communication and wireless power transmission papers.

In information collection stage, the UAV receives the data from the sensors of the $n$-\textit{th} group using $M-1$ antennas and sends the wireless power to the sensors of the $(n+1)$-\textit{th} group with the other single antenna. Assuming that the first antenna in the UAV antenna array is installed to send energy to the sensors and the other existing antennas are installed to receive information from the sensors, the uplink channel power gain from the $i$-\textit{th} sensor in the nth group to the $k$-\textit{th} antenna of the UAV and the downlink channel power gain from the $k$-\textit{th} antenna of the UAV to the $i$-\textit{th} sensor in the $(n+1)$-\textit{th} group are calculated by \eqref{eq:8} and \eqref{eq:9}, respectively. For ease of illustration, we consider all the channels of network have block fading. In this type of channels, the characteristics of the channel remain constant during a period of transmission in a block, but may change from one block to another.
\begin{equation}
	h_{ki}^{(n)}(t) = \frac{{{k_0}}}{{{{\left( {d_{ki}^{(n)}(t)} \right)}^2}}} = \frac{{{k_0}}}{{{{\left( {L_{ki}^{(n)}(t)} \right)}^2} + {A^2}}},\,\,\,k \in \{ 2,...,M\} 
	\label{eq:8}
\end{equation}

\begin{equation}
	g_{ki}^{(n + 1)}(t) = \frac{{{k_0}}}{{{{\left( {d_{ki}^{(n + 1)}(t)} \right)}^2}}} = \frac{{{k_0}}}{{{{\left( {L_{ki}^{(n + 1)}(t)} \right)}^2} + {A^2}}},\,\,\,\,k = 1
	\label{eq:9}
\end{equation}

where $k_0$ is the channel power gain at a distance of one meter from the sensor. In addition, we assume that the UAV is aware of the channel power gain, and receives the information of its next destination at the current location. This means the sensors of the $n$-\textit{th} group inform the UAV of the location of the $(n+1)$-\textit{th} group during the period of sending their information. As a consequence, in this 2D model, the UAV can be moved sequentially from the first group to the $N$-th group. It should be noted that the uplink channel is only used to send information from the sensor to the UAV, and the UAV antenna array is fixed on the desired group of sensors, so the gain of the uplink channel is not a function of time. This means:
\begin{equation}
	\begin{split}
		h_{ki}^{(n)} &  = \frac{{{k_0}}}{{{{\left( {d_{ki}^{(n)}} \right)}^2}}} = \frac{{{k_0}}}{{{{\left( {L_{ki}^{(n)}} \right)}^2} + {A^2}}} \\
		& = {\frac{{{k_0}}}{{{{({x_{n - 1}} - {x_i})}^2} + {{(\tilde{y} + (k - 1)\delta  - {y_i})}^2} + {A^2}}}} ,\,\,\,k \in \{ 2,...,M\}
	\end{split}
	\label{eq:10}
\end{equation}

\subsection{Energy Harvesting}
\begin{figure}
\centering{\includegraphics[scale=0.45]{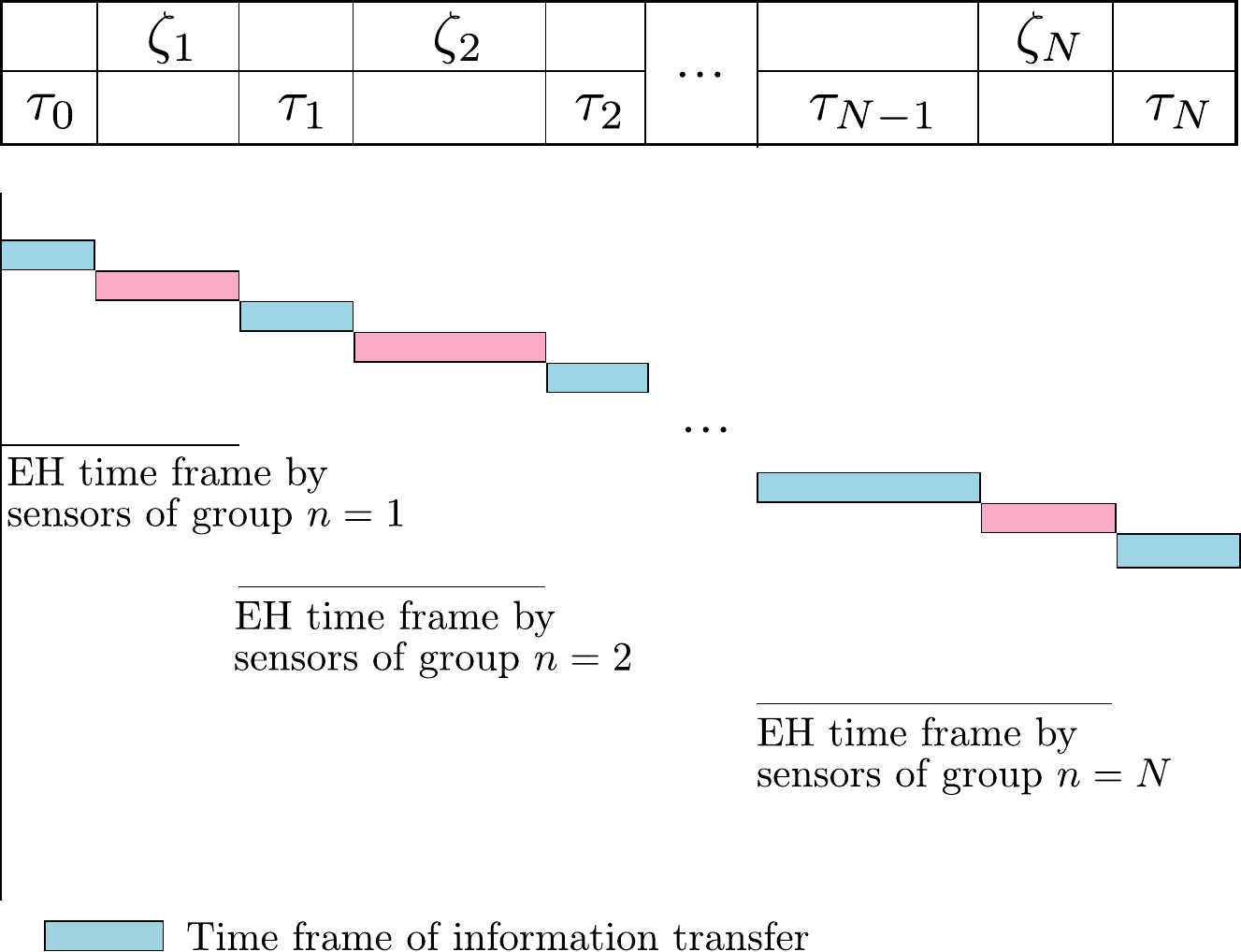}}
\caption{Time frame for information transmission and energy harvesting.}\label{Fig3}
\end{figure}

Using TDMA structure, we consider energy harvesting and information transmission through a transmission block in Figure~\ref{Fig3}, where the sensors in each group can send information only in the time slot, allocated to them. So, there is no interference between the information sent by the sensors of different groups. Also, in each block, the UAV broadcasts the energy signal to all users with a fixed transmission power, using one of its antennas. In contrast to frequency division multiple access (FDMA) and orthogonal frequency division multiple access (OFDMA), the TDMA architecture enables simple implementation of the WPT protocol in the sensors.

When the distance among the UAV and the sensors is less than the maximum communication distance to receive effective energy, the sensors in the desired group can receive energy. It should be noted that only the first antenna of the UAV performs the act of energy supply, i.e., $k=1$. Therefore, the total energy received by the $i$-\textit{th} sensor of the $n$-\textit{th} group can be calculated as follows:

\begin{equation}
	E_{ki}^{(n)}(t) = E_{ki}^{(n),hh}(t) + E_{ki}^{(n),hf}(t),\,\,\,\, \forall i \in \{1,2,\dots,K\},  \label{eq:11}
\end{equation}
where $E_{1i}^{(n),hf}(t)$ is the energy received by the $i$-\textit{th} sensor of the $n$-\textit{th} group from the UAV, when the UAV is flying from the $(n-1)$-\textit{th} group to the $n$-\textit{th} group. The $E_{1i}^{(n),hh}(t)$,  as the energy received by the $i$-\textit{th} sensor of the $n$-\textit{th} group from the UAV, when the UAV is hovering over the $n$-\textit{th} group, can be computing as \eqref{eq:12}.
\begin{equation}
	E_{1i}^{(n),hh}(t) = \eta_iP_tg_{1i}^{(n)}\tau_{n-1}, \label{eq:12}
\end{equation}
where $\eta_i \in \left(\left.0,1\right]\right. $
is the effective energy harvesting factor for the $i$-\textit{th} sensor. Here, a linear energy harvesting model is considered, since when the RF signals reach the ground sensors, the input power in the sensor is in the linear zone.

The operation of energy harvesting by the sensors during the UAV flight is depends on the different available air to ground channels according to the distance among the UAV and the sensors. Next, $E_{1i}^{(n),hf}(t)$ is equal to the following equation:

\begin{equation}
	E_{1i}^{(n),hf} = {\eta _i}\int\limits_0^{{\zeta _n}} {{P_t}g_{1i}^{(n)}(t)\mathrm{d}t = } {\eta _i}\int\limits_0^{{\zeta _n}} {{P_t}\frac{{{k_0}}}{{{{\left( {L_{1i}^{(n)}(t)} \right)}^2} + {A^2}}}\mathrm{d}t} \label{eq:13}
\end{equation}
According to the odd and even horizontal paths of the UAV movement in the network, the calculations of the energy harvesting operation will be as follows:

\begin{equation}
	E_{1i}^{(n),hf} =
	\begin{cases}
		\eta _i\int_0^{\zeta _n} {P_t\frac{k_0}{({v_n}t + {x_{n - 1}} - {x_i})^2 + {(\tilde{y} - y_i)}^2 + A^2}} \,\mathrm{d}t	& n \in \mathbb{N}_r,\, \textrm{r is odd} \\ 
		{\eta _i}\int\limits_0^{{\zeta _n}} {{P_t}\frac{{{k_0}}}{{{{( - {v_n}t + {x_{n - 1}} - {x_i})}^2} + {{(\tilde{y} - {y_i})}^2} + {A^2}}}} \,\mathrm{d}t		& n \in \mathbb{N}_r,\, \textrm{r is even}
	\end{cases}
	\label{eq:14}
\end{equation}
Finally, the total received energy $E_i^{(n)}$  can be rewritten as \eqref{eq:15} using \eqref{eq:12} and \eqref{eq:14}.
\begin{equation}
	E_i^{(n)} = {\eta _i}{P_t}{k_0}\left( {a_i^{(n)}{\tau _{n - 1}} + b_i^{(n)}{\zeta _n}} \right) \label{eq:15}
\end{equation}
where we have
\begin{equation}
	a_i^{(n)} = \frac{1}{{{{\left( {{x_n} - {x_i}} \right)}^2} + {{\left( {\tilde{y} - {y_i}} \right)}^2} + {A^2}}}  \label{eq:16}
\end{equation}
\begin{equation}
b_i^{(n)} =
	\begin{cases}
	{\frac{1}{{{D_n}\sqrt {{A^2} + {{\left( {\tilde{y} - {y_i}} \right)}^2}} }}\left[ {\arctan \left( {\frac{{{D_n} + {x_{n - 1}} - {x_i}}}{{\sqrt {{A^2} + {{\left( {\tilde{y} - {y_i}} \right)}^2}} }}} \right) - \arctan \left( {\frac{{{x_{n - 1}} - {x_i}}}{{\sqrt {{A^2} + {{\left( {\tilde{y} - {y_i}} \right)}^2}} }}} \right)} \right]},\!\!&\!\! n \in \mathbb{N}_r, \textrm{r is odd} \\ 
	{\frac{1}{{{D_n}\sqrt {{A^2} + {{\left( {\tilde{y} - {y_i}} \right)}^2}} }}\left[ {\arctan \left( {\frac{{{D_n} - {x_{n - 1}} + {x_i}}}{{\sqrt {{A^2} + {{\left( {\tilde{y} - {y_i}} \right)}^2}} }}} \right) - \arctan \left( {\frac{{ - {x_{n - 1}} + {x_i}}}{{\sqrt {{A^2} + {{\left( {\tilde{y} - {y_i}} \right)}^2}} }}} \right)} \right]},\!\!&\!\! n \in \mathbb{N}_r, \textrm{r is even}
	\end{cases}
\label{eq:17}
\end{equation}
It is assumed that the sensor does not have a battery and only uses a super capacitor for communication, so the entire energy received by the sensor is spent on transmitting information in the corresponding time interval. The instantaneous transfer rate per bandwidth unit for the $i$-\textit{th} sensor of the $n$-\textit{th} group is as follows:
\begin{equation}
	\begin{split}
		\left\{ R^{(n)} \right\}   & = \sum_{i \in {S_n}} {\sum_{k = 2}^M {R_{ki}^{(n)} } }  \\
		& \le \frac{1}{2}\log \left[ {1 + {\eta _i}{P_t}{k_0}\sum_{i \in {S_n}} {\sum_{k = 2}^M {\frac{{h_{ki}^{(n)}\left( {a_i^{(n)}{\tau _{n - 1}} + b_i^{(n)}{\zeta _n}} \right){{\left( {h_{ki}^{(n)}} \right)}^T}}}{{{\tau _n}{\sigma ^2}}}} } } \right] \\
		& =\frac{1}{2}\log\! \left[\! {1 \!+\! \frac{{{\gamma ^{(n)}}\left( {{\tau _{n - 1}}\sum\limits_{i \in {S_n}} {a_i^{(n)} \!+\! {\zeta _n}\sum\limits_{i \in {S_n}} {b_i^{(n)}} } } \right)}}{{{\tau _n}}}}\! \right],\! i \in \{1,2,\dots,K\},(\mathrm{Nats/Sec/Hz})
	\end{split}
	\label{eq:18}
\end{equation}
where $\gamma ^{(n)} = \frac{{{\eta _i}{P_t}{k_0}}}{{{\sigma ^2}}}\sum\limits_{i \in {S_n}} {\sum\limits_{k = 2}^M {h_{ki}^{(n)}{{\left( {h_{ki}^{(n)}} \right)}^T}}}  $, and $\sigma ^2$  is the noise power of the channel.
	
\section{Sum Throughput Maximization}\label{sec:STM}
In this section, we have considered an optimization problem for the distribution of the UAV flight and hovering time in order to achieve the maximum amount of the sum throughput of an IoT network equipped with a UAV. The optimal problem can be solved by using mutual coupling conditions in convex optimization problems and generalizing the relations used in \cite{17}. At the same time, a relationship is established between the flight time of the UAV and the time it hovers over the groups. We have considered the initial hovering time of the UAV on the flight starting point to be as small as possible and we have set the UAV's flight speed to the maximum state. Then, the other hovering and flying times of the UAV on the desired groups are obtained according to the algorithm~\ref{alg1}. We will see that for a feasible problem, $\sum\limits_{k = 2}^M {\sum\limits_{i = 1}^K {L_{ki}^{(n)} \le {V_{\max }}T} }$ holds, so a feasible value for the distribution of flight and hovering time can be obtained, when the UAV moves from the first group to the last group of sensors in the network in the time period $T$. By representing the non-negative duration as $\tau_n$  and $\zeta_n$  with the vectors ${\boldsymbol{\tau }} = {\{ {\tau _0},...,{\tau _N}\} ^T}$ and ${\boldsymbol{\zeta }} = {\{ {\zeta _1},...,{\zeta _N}\} ^T}$, the STM problem is defined as follows:

\begin{subequations}
	\begin{align}
			&{\max _{\tau ,\zeta }}\sum\limits_{n = 1}^N {{\tau _n}{R_n}} , \label{19a}\\
			{\mathrm{s.t.}} &\,\, {\tau_0 \ge 0, \forall n\in\aleph},\label{19b}\\
			& {V_{\mathrm{max}} \ge \frac{D_n}{\zeta_n}, \forall n \in \aleph,}\label{19c}\\
			&\sum_{n=0}^N {\tau_n} + \sum_{n=1}^N {\zeta_n} \le T. \label{19d}
	\end{align}
\end{subequations}
The constraint \eqref{19b} is due to the non-negativity property of the hover time, \eqref{19c} is the considered limit for the maximum speed of the UAV and \eqref{19d} is the time limit for a complete operation. It is considered, that all sensors have a minimum amount of information to send. To solve the optimization problem raised in \eqref{19a}-\eqref{19d}, according to its constraints, we have used the proof in \cite{14}. The Throughput function for the $n$-\textit{th} group of sensors can be written as ${\varphi _n}(\tau ,\zeta ) = {\tau _n}{R_n},\forall n \in \aleph $, which is a concave function with respect to the flight and hovering time of the UAV. To prove that \eqref{19a}-\eqref{19d} is a convex optimization problem, we first give the proposition~\ref{prop1}.

\begin{prop}\label{prop1}
If we define the throughput function of group $n$ as ${H_n}(\tau ,\zeta ) \buildrel \Delta \over = {\tau _n}{R_n}$, then the ${H_n}$ is a concave function of $\tau$ and $\zeta$.
\end{prop}

All constraints of this problem are also affine. Therefore, the STM is a convex optimization problem, and can be solved using convex optimization tools such as CVX \cite{cvx}. However, due to the high complexity in this type of calculations with the increase in the number of the sensors in the network, another algorithm has been used to solve the problem. To obtain the optimal flight and hovering time of the UAV for the STM problem, we introduce two variables $F_1$ and $F_2$:
	
\begin{equation}
		{F_1} \buildrel \Delta \over = \left[ {T - \sum\limits_{n = 2}^{N - 1} {\left( {\sum\limits_{c = n + 1}^N {\prod\limits_{z = n + 1}^c {\frac{{{a_z}}}{{{f_z}}} + 1 + \frac{{{f_n}}}{{{b_n}}}} } } \right){b_n}\frac{{{\zeta _n}}}{{{f_n}}} - \left( {1 + \frac{{{f_N}}}{{{b_N}}}} \right){b_N}\frac{{{\zeta _N}}}{{{f_N}}}} } \right]\prod\limits_{n = 1}^N {{f_n}} \label{eq:20}
\end{equation}
	and
\begin{equation}
		{F_2} \buildrel \Delta \over = \prod\limits_{n = 1}^N {{f_n} + {b_1}\left( {\prod\limits_{n = 2}^N {{f_n} + \prod\limits_{n = 2}^N {{a_n} + \sum\limits_{n = 2}^{N - 1} {\prod\limits_{c = 2}^n {{a_c}\prod\limits_{z = n + 1}^N {{f_z}} } } } } } \right)} \label{eq:21}
\end{equation}
	where $f_n$ is obtained as coupling variables as follows:
\begin{equation}
		{f_N} = \frac{1}{{{\gamma _N}}}\left( {\frac{{{\gamma _N}{b_N} - 1}}{{\mathcal{W}\left( {\left( {{\gamma _N}{b_N} - 1} \right){{\exp }^{ - {\mu _N} - 1}}} \right)}} - 1} \right), \label{eq:22}
\end{equation}
\begin{equation}
		{f_n} = \frac{1}{{{\gamma _N}}}\!\left(\! {\frac{1}{{ - \mathcal{W}\left( { - \exp \left( {\frac{{{\gamma _{n + 1}}{a_{n + 1}}}}{{1 + {\gamma _{n + 1}}{f_{n + 1}}}} - \frac{{{\gamma _N}{b_N}}}{{1 + {\gamma _N}{f_N}}} - \left( {{\mu _N} + 1} \right)} \right)} \right)}} \!-\! 1}\! \right)\!,\!\forall n \in \{N-1,\dots,1\}\label{eq:23}
\end{equation}
where $\mathcal{W}(.)$  is known as the Lambert W function \cite{24}. The $\mu_N$ is a non-negative dual Lagrange variable, which can be calculated as ${\mu _N} = \mathrm{Root}\left( {\frac{{{\gamma _1}{b_1}}}{{{Y_1}}} - \frac{{{\gamma _N}{b_N}}}{{{Y_N}}} = {\mu _N}} \right)$  by replacing $n=N$ in \eqref{19c}.
	
Here we have ${Y_N} \buildrel \Delta \over = \frac{{{\gamma _N}{b_N} - 1}}{{\mathcal{W}\left( {\left( {{\gamma _N}{b_N} - 1} \right){{\exp }^{ - {\mu _N} - 1}}} \right)}}$, ${Y_1} = {G_1}\left( {...{\rm{ }}{{\rm{G}}_{N - 1}}({Y_N}){\rm{ }}...} \right)$ and ${G_n}({Y_n}) \buildrel \Delta \over = \frac{1}{{ - \mathcal{W}\left( { - \exp \left( {\frac{{{\gamma _n}{a_n}}}{{{Y_n}}} - \frac{{{\gamma _N}{b_N}}}{{{Y_N}}} - \left( {{\mu _N} + 1} \right)} \right)} \right)}}$.
	
Also, Root(.) means to find the root of the desired equation. Finally, the optimal fly and hovering time of the UAV for all groups is obtained using Algorithm~\ref{alg1}.

\begin{algorithm}[H]
\caption{\textsc{Sum Throughput Maximization Problem}}  
\label{alg1} 
\begin{algorithmic}[1]
\State Find ${\mu _N} = \mathrm{Root}\left( {\frac{{{\gamma _1}{b_1}}}{{{Y_1}}} - \frac{{{\gamma _N}{b_N}}}{{{Y_N}}} = {\mu _N}} \right)$ 
\State Find $f_N$ using \eqref{eq:22} 
\State Calculate $f_n$ from $n=N-1$ to $n=1$ using \eqref{eq:23}
\State Calculate $\zeta_1=\frac{F_1}{F_2}$ using \eqref{eq:20}-\eqref{eq:21} 
\State Set $\tau_1=\frac{b_1\zeta_1}{f_1}$
\For {$n=2$ to $N$} 
\State Find: $\zeta_n=\frac{D_n}{V_{\mathrm{max}}}$  
\State Calculate: ${\tau _n} = \frac{{{a_n}{\tau _{n - 1}} + {b_n}{\zeta _n}}}{{{f_n}}}$   
\EndFor
\State \Return {$\boldsymbol{\tau}=[\tau_0,\dots,\tau_N]^T$}
\State \Return {$\boldsymbol{\zeta}=[\zeta_1,\dots,\zeta_N]^T$}  	
\end{algorithmic}
\end{algorithm}

\section{Total Time Minimization} \label{sec:TTM}
In this part, we have assumed that each sensor has the minimum amount of information, $I_n$, to send to the UAV. Based on this, an optimization problem is created to minimize the flight time of the UAV and the time it hovers over the desired groups in the form of $\sum_{n=0}^N \tau_n + \sum_{n=1}^N \zeta_n$. Finally, the problem of minimizing the total time of each operation is defined as follows:
		\begin{subequations}
			\begin{align}
				&{\min _{\tau ,\zeta }}{\sum_{n = 0}^N {\tau _n}} + {\sum_{n = 1}^N {\zeta _n}}, \label{24a}\\
				{\mathrm{s.t.}} \,&\,\, {\tau_n R_n \ge I_n, \forall n\in\aleph},\label{24b}\\
				& {\tau_0 \ge 0,\,\,\, \tau_n\ge 0, \forall n\in\aleph},\label{24c}\\
				& {V_{\mathrm{max}} \ge \frac{D_n}{\zeta_n}, \forall n \in \aleph,}\label{24d}
			\end{align}
		\end{subequations}
The equation \eqref{24b} guarantees that there is the minimum information required to be sent from the sensor to the UAV. According to the TTM problem and its constraints, the optimal flight and hovering time of the UAV for each group of sensors is achieved.
		\begin{equation}
			{\tau _n} = \frac{{{I_n}}}{{\mathcal{W}\left( {\frac{{\left( {1 - \frac{{{a_{n + 1}}}}{{{b_{n + 1}}}}} \right){\gamma _n}{b_n} - 1}}{{\exp (1)}}} \right) + 1}},\,\,\, \forall n \in\{1,\dots,N-1\} \label{eq:25}
		\end{equation} 
		\begin{equation}
			{\tau _N} = \frac{{{I_N}}}{{\mathcal{W}\left( {\frac{{{\gamma _n}{b_n} - 1}}{{\exp (1)}}} \right) + 1}}, \label{eq:26}
		\end{equation}
		\begin{equation}
			{\zeta _n} = \frac{{\frac{{{\tau _n}}}{{{\gamma _n}}}\left( {\exp \left( {\frac{{{I_n}}}{{{\tau _n}}} - 1} \right)} \right) - {a_n}{\tau _{n - 1}}}}{{{b_n}}},\,\,\forall n \in\{2,\dots,N\}, \label{eq:27}
		\end{equation}
		\begin{equation}
			{\zeta _1} = \frac{{\frac{{{\tau _1}}}{{{\gamma _1}}}\left( {\exp \left( {\frac{{{I_1}}}{{{\tau _1}}} - 1} \right)} \right)}}{{{b_1}}}. \label{eq:28}
		\end{equation}
		
By implementing Algorithm~\ref{alg2}, the complexity created through convex optimization calculation tools can be reduced. Also, you can use some values, which have already been obtained in Algorithm~\ref{alg1}.
		
\begin{algorithm}[H]
\caption{\textsc{Total Time Minimization problem}.}  
\label{alg2} 
\begin{algorithmic}[1]
\State Find $\tau_N$ using \eqref{eq:26}
\For {$n=N-1$ to $n=1$} 
\State {Calculate $\tau_n$ using \eqref{eq:25}}
\State {Calculate ${\zeta _{n + 1}} = \max \left( {\frac{{\frac{{{\tau _{n + 1}}}}{{{\gamma _{n + 1}}}}\left( {\exp \left( {\frac{{{I_{n + 1}}}}{{{\tau _{n + 1}}}} - 1} \right)} \right) - {a_{n + 1}}{\tau _n}}}{{{b_{n + 1}}}},\frac{{{D_{n + 1}}}}{{{V_{\max }}}}} \right)$}
\EndFor
\State Calculate ${\zeta _1} = \max \left( {\frac{{\frac{{{\tau _1}}}{{{\gamma _1}}}\left( {\exp \left( {\frac{{{I_1}}}{{{\tau _1}}} - 1} \right)} \right)}}{{{b_1}}},\frac{{{D_1}}}{{{V_{\max }}}}} \right)$
\State \Return {$\boldsymbol{\tau}=[\tau_0,\dots,\tau_N]^T$}
\State \Return {$\boldsymbol{\zeta}=[\zeta_1,\dots,\zeta_N]^T$}  
\end{algorithmic}
\end{algorithm}
		
\section{Numerical Simulations}\label{sec:simulations}
In this section, we analyze the efficiency of our network scheme and compare it with the results of the hover-and-fly energy harvesting (HF-EH) network scheme in \cite{14}. All simulations are done using MATLAB 9.8.0 (R2020a) software. First, we investigate the effect of changes in the transmission power of the UAV, the number of groups of the sensors and, consequently, the number of sensors in the network in the STM problem. Next, we discuss the effect of changes in the transmission power of the UAV as well as changes in its maximum horizontal speed in the TTM problem.

All results are averaged over 100000 independent channel realizations in the Monte-Carlo experiments. In the simulations, we determine the distances among the groups of sensors are random variables, uniformly distributed in the interval $D_n=[20,30)$ m. Then, we consider the frequency of the network as $f_0=3\,\,\mathrm{GHz}$, the number of antennas as $M=3$, and the distance among the antennas in the UAV antenna array as $0.1$ m. In this work, we limit the maximum speed experienced by the UAV during its flight as $V_{\mathrm{max}}=10\, \mathrm{m/sec}$. The time to complete each UAV operation in the target network is fixed at $T=1000$ sec. The simulation parameters are given in Table~\ref{table}.
		
		\begin{table}[H]
			\centering
			\caption{Simulation Parameters.}
			\vspace*{0.3cm}
			\begin{tabular}{c|c|c}
				\hline
				\hline
				Parameter & Desciption & Value  \\
				\hline
				\hline
				$k_0$ & Channel power gain at a distance of 1 meter & $-30$ dB  \\
				\hline
				$\sigma^2$ & Noise power & $-70$ dBm  \\
				\hline
				$A$ &  Height of the UAV & $10$ m  \\
				\hline
				$\tilde{y}$ & Vertical coordinates of the UAV reference antenna & $[0,5)$ m  \\
				\hline
				${\eta_i}$ & Effective amount of harvested energy & $0.5$  \\
				\hline
				${M}$ & Number of Antennas & $3$  \\
				\hline
				$f_0$ & Frequency of the network & $f_0=3\,\,\mathrm{GHz}$  \\
				\hline
				$\delta$ & Distance among the antennas in the UAV antenna array & $0.1$ m \\
				\hline
				$V_{\mathrm{max}}$ & Maximum UAV speed & $10\, \mathrm{m/sec}$ \\
				\hline
				$T$ & Total time of each UAV operation & $1000$ \\
				\hline
				$D_n$ & Distances among the groups of sensors & $[20,30)$ m \\
				\hline
				$K$ & Number of sensors (except Figure~\ref{Fig5}) & $20$ \\
				\hline
				$N$ & Number of sensor groups (except Figure~\ref{Fig5}) & $4$ \\
				\hline

			\end{tabular}
			
			\label{table}
		\end{table}

\subsection{The STM Problem}
\begin{figure}[H]
	\centering{\includegraphics[scale=0.75]{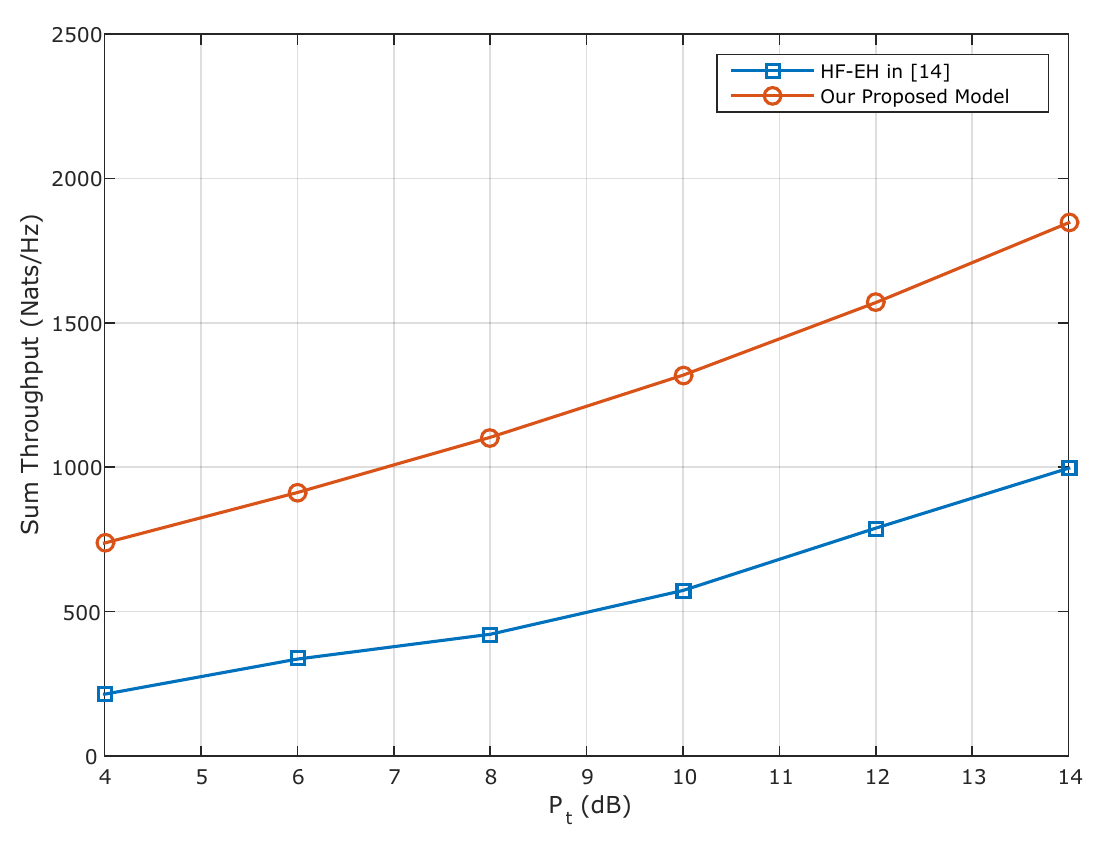}}
	\caption{The effect of changes in transmission power $P_t$, on the sum throughput.}\label{Fig4}
\end{figure}

In Figure~\ref{Fig4}, according to the results obtained in this research, we evaluate the effect of changes in $P_t$ on the sum throughput. We divide $K=20$ sensors into $N=4$ groups. As can be seen, the final operational capacity of both plans increases with the increase of $P_t$ of the UAV, since more $P_t$ allows the sensors to receive more energy to send data to the UAV and the rate of sending information increases. According to the diagram, it is clear that our proposed scheme, provide better performance than the HF-EH scheme, and we see a double growth of the values in the proposed network diagram. 
Also, when $P_t$ increases, the distance between two graphs also increases. This shows the importance of time allocation as the $P_t$ increases.
Using the MIMO antenna array in the UAV, and collecting data simultaneously from several sensors at each stop, allows us to increase the rate of sending information from the ground sensors to the UAV.
		
\begin{figure}[H]
	\centering{\includegraphics[scale=0.75]{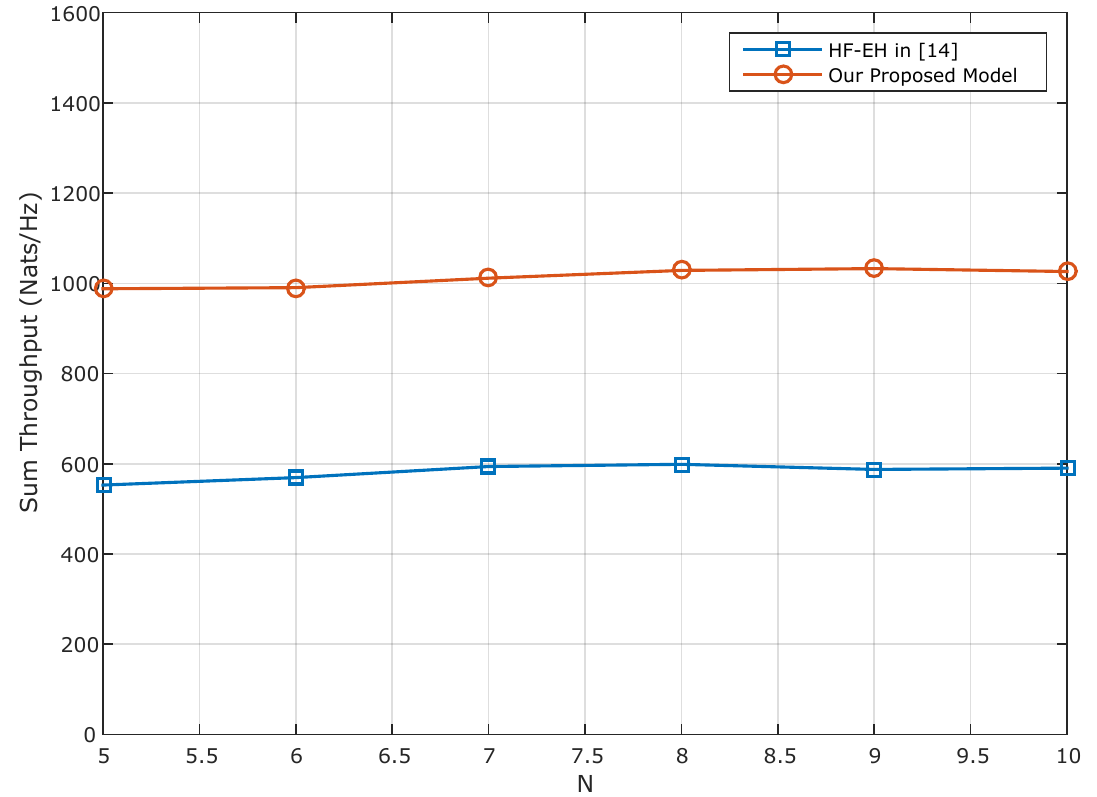}}
	\caption{The effect of changes in the number of  sensor groups $N$ on the sum throughput, with $P_t = 4$ dB.}\label{Fig5}
\end{figure}
		
In Figure~\ref{Fig5}, we evaluate the effect of changes in the number of group of sensors on the final throughput. We consider the transmission constant power $P_t=4$ dB and draw the plots for $N=5$ to $N=10$ groups. 
In both designs, as the number of groups increases, the final throughput also increases imperceptibly, until the graphs reach to a stable level. According to the values obtained in the numerical simulations, presented in the Table~\ref{table2}, our proposed scheme provides better performance than the HF-EH scheme by approximately 75\%.

\begin{table}[H]
	\centering
	\caption{Performance improvement rate of sum throughput under different values of $N$.}
	\vspace*{0.3cm}
	\begin{tabularx}{0.8\textwidth} { 
			| c 
			| >{\centering\arraybackslash}X 
			| >{\centering\arraybackslash}X
			| >{\centering\arraybackslash}X |}

		\hline
		\hline
		$N$ & {Sum throughput of our scheme} & {Sum throughput of the HF-EH scheme} &  {Performance improvement rate}  \\
		
		\hline
		\hline
		
		6 & {990} & {569} &  {74\% } \\
		\hline
		9 & {1033} & {587} & {76\%} \\
		\hline

	\end{tabularx}
	
	\label{table2}
\end{table}

Increasing the number of sensors in the network increases the number of times the UAV hovers and flies in the network; however, due to the constant time of each operation, the time of each hover and flight also decreases. As a result, in the HF-EH design, less time is provided to the UAV to collect information from each sensor. The grouping of sensors and the implementation of MIMO technology in our proposed scheme have caused this problem to be solved and to achieve better results.
		\begin{figure}[H]
			\centering{\includegraphics[scale=0.75]{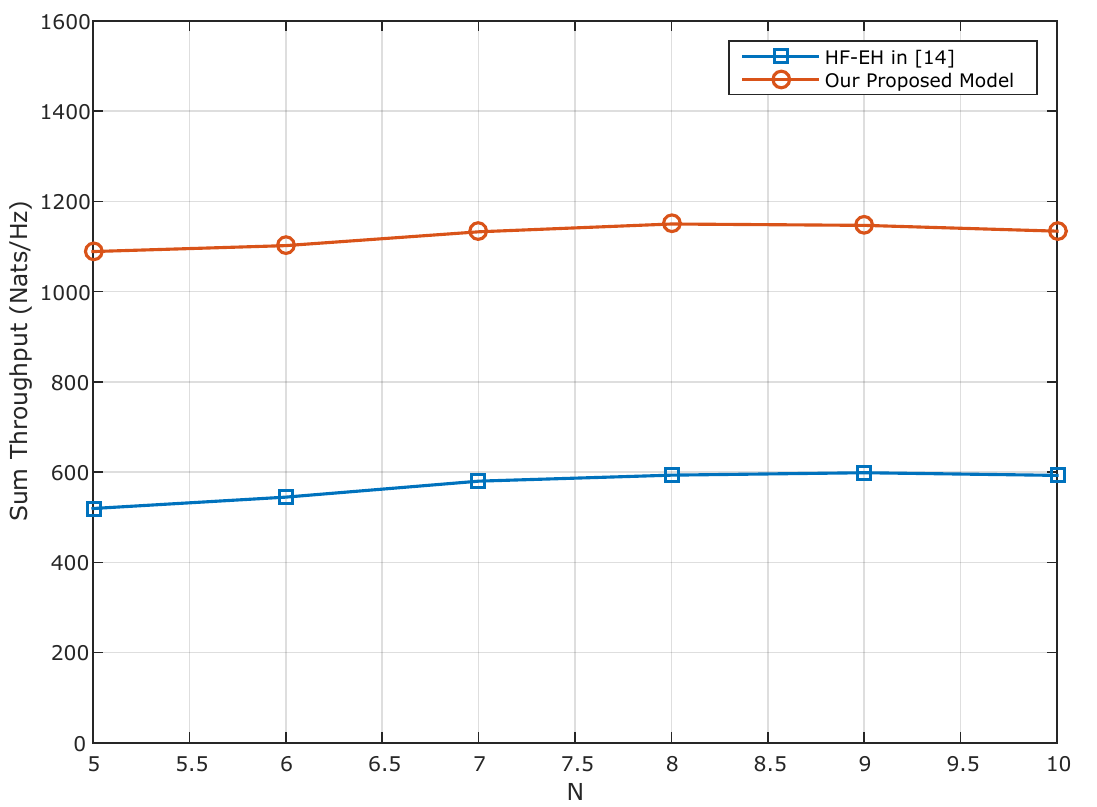}}
			\caption{The effect of changes in the number of sensor groups, $N$, on the sum throughput, with $P_t = 8$ dB.}\label{Fig6}
		\end{figure}
		
The positive effect of increasing the transmission power from $P_t = 4$ dB to $P_t = 8$ dB is also evident by comparing two figures~\ref{Fig5} and ~\ref{Fig6}. Our proposed scheme has 92\% better performance than the HF-EH scheme.
		
		\begin{figure}
			\centering{\includegraphics[scale=0.75]{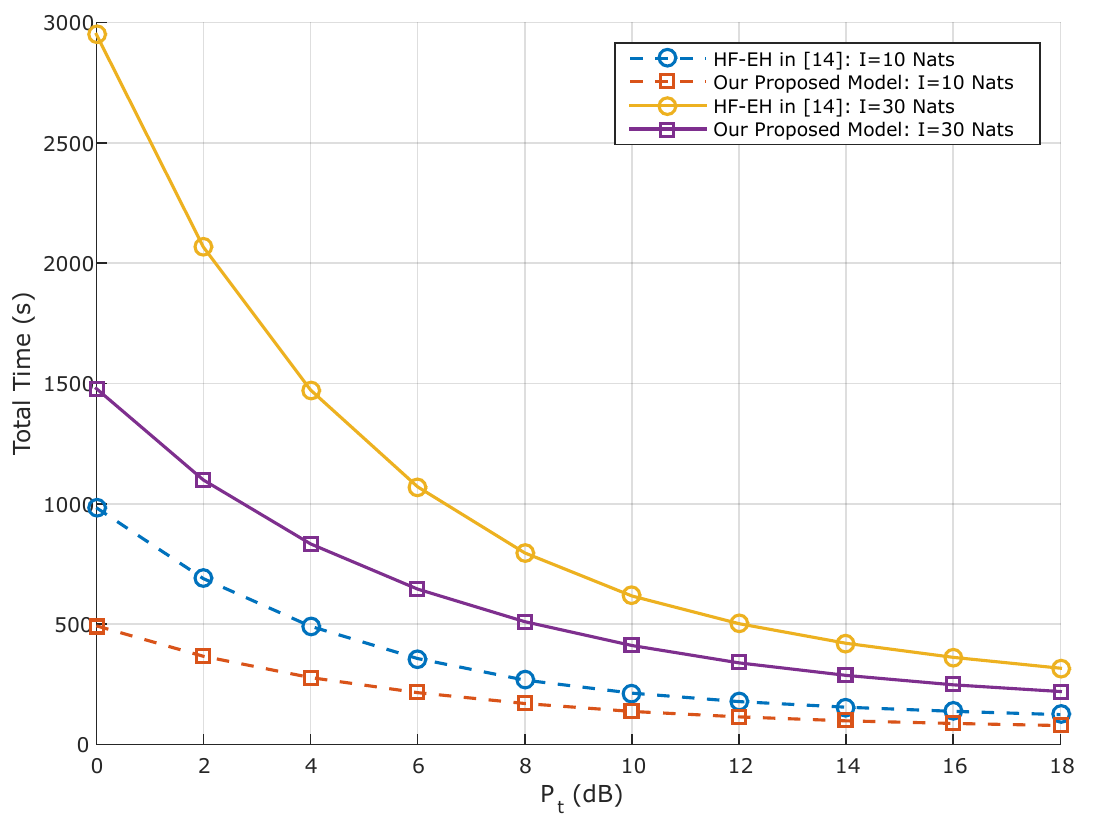}}
			\caption{The effect of transmission power $P_t$ changes on the total time of each operation $T$.}\label{Fig7}
		\end{figure}
\subsection{The TTM Problem}
Figure~\ref{Fig7} are depicted with the aim of analyzing the effect of changes in the transmission power on the total time of each operation. Here, the minimum amount of information, sent to the UAV by each sensor, is equal. To check the effect of the minimum amount of information on the system performance, we consider two values: $I=10$ Nats and $I=30$ Nats. Based on the Figure~\ref{Fig7}, the amount of time required to perform each operation in both our proposed scheme and the HF-EH scheme is reduced by increasing the transmission power.
Therefore, if the amount of transmission power increases, the time to transfer information from the sensor to the UAV will be shorter. As a result, the total amount of time, the UAV flies and hovers over the sensors, is reduced and the total time of each operation is minimized.
In equal conditions, due to the sending information from the sensors as a group and collecting information by the antenna array in the UAV, our proposed scheme requires less time than the HF-EH scheme. In fact, the time required for each operation is halved (For example, when $I=30$ Nats, and $P_t=2$ dB, the total time of our proposed scheme is 1007 s, and the total time of the HF-EH scheme is 2015 s). 
However, with the increase of transmission power, due to the limitations specified in the algorithm~\ref{alg2}, the slope of all graphs decreases, until it reaches a stable level. As indicated by \eqref{24d}, all IoT systems require a minimum total flight time for the UAV. As can be seen, by increasing the minimum amount of information sent, the required time in both schemes increases. So, both schemes have equal tolerance against changes in the amount of information loaded in the network. However, according to \eqref{24b}, we reach a stable level. As expected, it is clear that sending more information requires more time.
		\begin{figure}
			\centering{\includegraphics[scale=0.75]{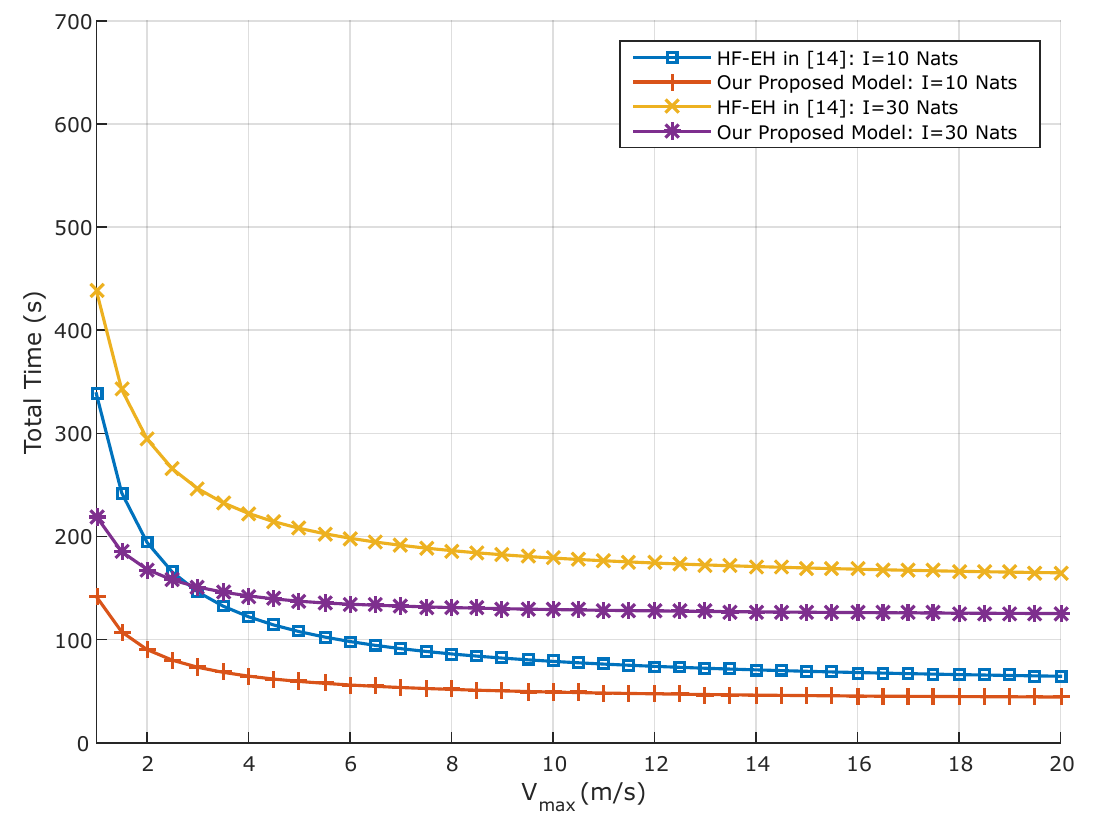}}
			\caption{The effect of changes in the maximum flight speed of the UAV, $V_{\mathrm{max}}$, on the total time of each operation $T$, with $P_t=30$ dB.}\label{Fig8}
		\end{figure}
		
Figure~\ref{Fig8}, shows the effect of the maximum UAV flight speed changes among the groups on the total time of each operation. It is obvious that as the speed of the UAV increases, the flight time of the UAV decreases, and then the total time of each operation will also reach to its minimum level. Further, according to the limitations mentioned in \eqref{24a}-\eqref{24d}, the plot reaches a constant level. Similar to Figure~\ref{Fig7}, it can be concluded that sending more information requires more time; however, when more information is sent, the distance between the plots of the two schemes will be greater. This means that our proposed scheme shows better performance, when we need to send more information from the sensors to the UAV. 
		
\begin{figure}
			\centering{\includegraphics[scale=0.75]{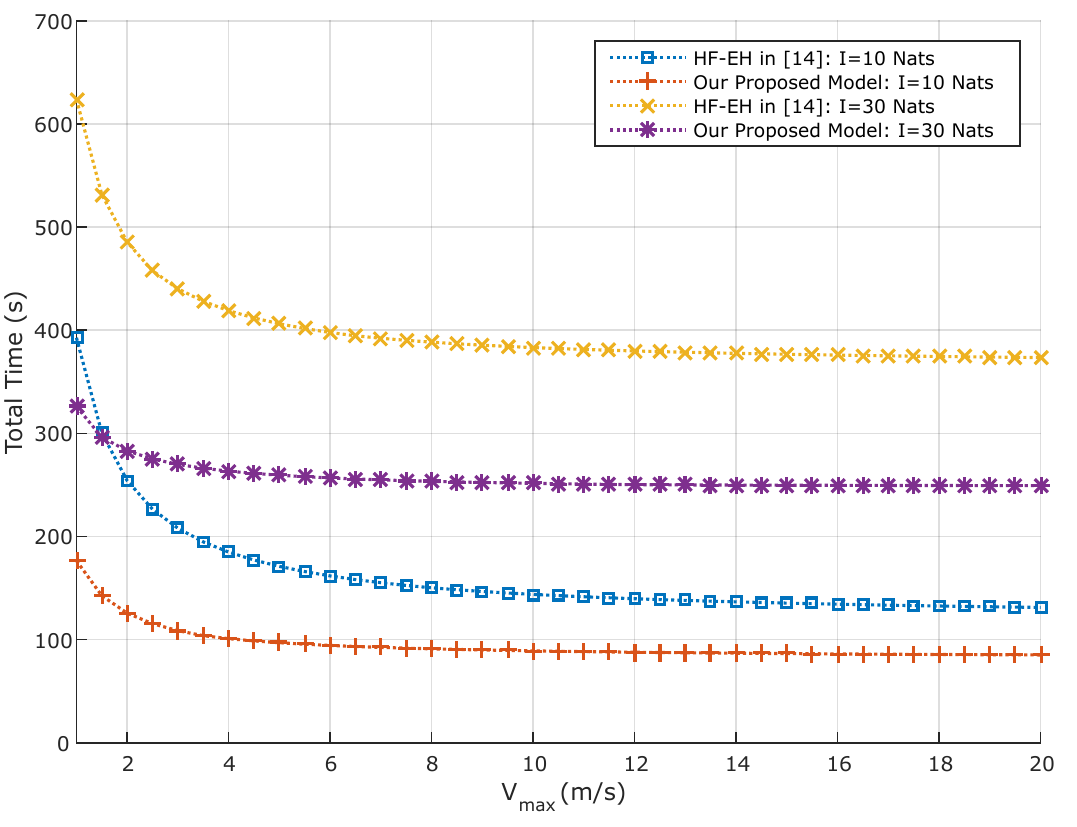}}
			\caption{The effect of changes in the maximum flight speed of the UAV, $V_{\mathrm{max}}$, on the total time of each operation $T$, with $P_t=15$ dB.}\label{Fig9}
		\end{figure}

Figure~\ref{Fig9} is drawn with the same values as Figure~\ref{Fig8}, while we have chosen the transmission power $P_t=15$ dB. A decrease in the transmission power component increases the minimum total time of each operation in all graphs. Increasing the minimum amount of information sent and simultaneously increasing the transmission power creates a greater distance in the graphs, which means that less time is needed in the network.
		
\section{Conclusion}\label{sec:Conclusion}
In this research, we propose a novel scheme for the rotary-wing UAV-enabled full-duplex wireless-powered IoT network, using the antenna array in the UAV. In this network, $K$ single antenna ground users are sparsely distributed. With the help of its antennas, the UAV provides wireless energy to the ground users and collects information from them. In contrast, sensors receive energy from the HAP and send their information to it. Due to the limitations in time resources, we used effective methods to manage time, and optimize it during operations. For this reason, we considered the sensors in the form of $N$ groups of sensors so that the UAV equipped with MIMO antenna array technology can serve the sensors as a group in the total time $T$.
By using the TDMA scheme to receive information from the users, we have implemented a simple WPT technology, i.e., each group of sensors receives energy from the UAV, when the UAV is hovering over the previous sensor group, and also when the UAV flies over the previous sensor group to the current sensor group. Each group of sensors sends its information to the UAV, when the UAV is hovering over it. Under this design, the sum throughput maximization (STM) problem and the total time minimization (TTM) problem are two optimization problems, which have been investigated in this research.
In the STM problem, by making changes in the transmission power of the UAV, we checked its effect on the network throughput and observed that with the increase in the transmission power, the network throughput will also increase. Then, we increased the number of groups of sensors, and observed that the final operational capacity of the network also increased at the same time. Also, instead of collecting information from a single sensor, the UAV can collect information from several sensors in a group at the same time. This made optimal use of the hovering time and thus increased the data transmission rate.
In the TTM problem, we increased the transmission power. In this case, the amount of time needed to perform each operation was also reduced to a stable level. The same result was obtained by increasing the maximum UAV flight speed.
In our proposed design, we have considered the total time of each operation in the STM problem fixed, while this time can be adjusted according to the amount of energy consumed by the UAV. We have also investigated the effect of a UAV equipped with an antenna array. Now it is possible to use several UAVs with the same feature in networks, which have a larger scale to increase the level of service coverage. The channel model used in this work is the quasi-static block fading channel. In order to use this network in the real world and model it accurately, the channel model can be considered dynamic and fast fading, and based on the characteristics of these types of channels, the problem can be analyzed. The energy considered in this work is only spent on flying and hovering the UAV, while servicing the sensors. However, receiving, storing and processing the information received in the UAV requires energy, which is not considered in current research, and is a part of our future work.

\appendix

\section{PROOF OF PROPOSITION 1}\label{appendix}
\begin{proof}
	According to the results obtained in \cite{22}, we know that if  $f:R^n\rightarrow R$ is a function, then the perspective function \cite{23} of $f$ will be a function such as $g:R^{n+1}\rightarrow R$, which is defined as $g(x,t)=tf(x/t)$  with domain $g={(x,t)|x/t \in \mathrm{dom} f,t>0}$ (dom refers to the domain of the function). Further, knowing that the perspective operation preserves the convexity, since $f$ is a concave function, so the function $g$ is also concave. Thus, as the following function is concave in $R_{++}^2$,
	\begin{equation}
		f({\tau _{n - 1}},{\zeta _n}) = \frac{1}{2}\log \left[ {1 + {\gamma ^{(n)}}\left( {{\tau _{n - 1}}\sum\limits_{i \in {S_n}} {a_i^{(n)} + } {\zeta _n}\sum\limits_{i \in {S_n}} {b_i^{(n)}} } \right)} \right],{\rm{  }}
	\end{equation}
Therefore, the perspective function of $f$, defined as:
	\begin{equation}
		g\left( {{\tau _{n - 1}},{\zeta _n},{\tau _n}} \right) = \frac{1}{2}{\tau _n}\log \left[ {1 + \frac{{{\gamma ^{(n)}}\left( {{\tau _{n - 1}}\sum\limits_{i \in {S_n}} {a_i^{(n)} + } {\zeta _n}\sum\limits_{i \in {S_n}} {b_i^{(n)}} } \right)}}{{{\tau _n}}}} \right],{\rm{ }}
	\end{equation}
	is concave in $R_{++}^3$. Also, the non-negative weighted sum of a concave function is always concave. Therefore, \eqref{19a}-\eqref{19d} is a convex optimization problem.
\end{proof}

\bibliographystyle{IEEEtran} 
\bibliography{references}

\end{document}